\documentclass[aps,twocolumn,prl,showpacs,multicol,amsmath,amssymb]{revtex4}
\usepackage[dvips]{graphicx}

\newcommand{\be}{\begin{equation}}
\newcommand{\ee}{\end{equation}}

\newcommand{\bea}{\begin{eqnarray}}
\newcommand{\eea}{\end{eqnarray}}
\newcommand{\bd}{\begin{displaymath}}
\newcommand{\ed}{\end{displaymath}}
\newcommand{\ba}{\begin{array}}
\newcommand{\ea}{\end{array}}
\newcommand{\bi}{\begin{itemize}}
\newcommand{\ei}{\end{itemize}}
\newcommand{\bc}{\begin{center}}
\newcommand{\ec}{\end{center}}
\newcommand{\bfl}{\begin{flushleft}}
\newcommand{\efl}{\end{flushleft}}
\newcommand{\bfr}{\begin{flushright}}
\newcommand{\efr}{\end{flushright}}

\def\6{\partial}

\def\={\!\!\!&=&\!\!\!}
\def\+{\!\!\!&&\!\!\!+~}
\def\-{\!\!\!&&\!\!\!-~}

\begin{document}

\title[]{Quasiparticle interference in the spin-density wave phase of iron-based superconductors
%probed in STM experiments
}
 \author {J. Knolle$^{1}$, I. Eremin$^{2}$, A. Akbari$^{2}$, and R. Moessner$^{1}$}
 \affiliation{
 $^{1}$Max Planck Institute for the Physics of Complex Systems, D-01187 Dresden, Germany}
 \affiliation {$^2$Institut f\"ur Theoretische Physik III, Ruhr-Universit\"at Bochum, D-44801 Bochum, Germany}
% \affiliation {$^3$Code 6393, Naval Research Laboratory, Washington, DC 20375, USA}

%
\begin{abstract}
We propose an explanation for the electronic nematic state observed recently in parent
iron-based superconductors [T.-M. Chuang {\it et al.,} Science {\bf 327}, 181 (2010)]. We argue that the quasi-one-dimensional nanostructure identified in the quasiparticle interference (QPI) is a consequence of the interplay of the magnetic $(\pi,0)$ spin-density wave (SDW) order with the underlying electronic structure. We show that the evolution of the QPI peaks largely reflects quasiparticle scattering between electronic bands involved in the SDW formation. Because of the ellipticity of the electron pocket and the fact that only one of the electron pockets is involved in the SDW, the resulting QPI has a pronounced one-dimensional structure. We further predict that the QPI crosses over to two-dimensionality on an energy scale, set by the SDW gap, which we estimate from neutron scattering data to be  around 90 meV.
\end{abstract}

\date{\today}


\maketitle

{\it Introduction.}
One of the key challenges in condensed matter physics is to understand the nature of the many-body states which manifest themselves in experimentally observable anomalous properties. Such states may exhibit subtle, or even entirely novel, forms of static or fluctuating order.
One of the most recent examples is the nematic electronic structure observed by means of spectroscopic imaging-scanning tunneling microscopy (SI-STM) experiments in Ca(Fe$_{1-x}$Co$_x$)$_2$As$_2$, parent material of iron-based superconductors\cite{chuang}. Given a certain
similarity between the phase diagrams of iron-based and cuprate superconductors\cite{kamihara08} -- both contain an antiferromagnetic phase at small, and a
superconducting phase at larger, dopings -- and in view of the checkerboard electronic pattern observed earlier in the cuprates\cite{hanaguri}, this experiment also refocuses attention on a  possible  role of quasi-one-dimensional physics. Indeed, the possibility of nematic order arising from orbital physics in such quasi-two-dimensional electronic systems has already been discussed in Ref.\cite{wu}.

Despite such similarities, there are considerable differences in the normal state electronic structure of the respective parent compounds. The iron-based compound
exhibits two circular hole pockets of unequal size, centered around the $\Gamma$ point $(0,0)$,  and two elliptic electron pockets
centered at the $(0, \pm \pi)$ and $(\pm \pi,0)$  points of the
unfolded Brillouin zone (UBZ, based on the $Fe$-lattice)~\cite{LDA,kaminski,coldea}. Electron and hole bands are significantly nested, {\it i.e.} $\varepsilon_{\bf k}^{h} \simeq -\varepsilon_{\bf k+Q_i}^{e}$ where ${\bf Q}_i$ is either ${\bf Q}_1 = (0,\pi)$ or ${\bf Q}_2=(\pi,0)$.
As nesting enhances SDW instabilities, several researchers have argued for an itinerant description of the magnetism in those compounds based at least partly on nesting~\cite{Tesanovic}. Also, in this picture the specific selection of a $(0,\pi)$ or $(\pi,0)$ magnetic order, as well as the anisotropy of the spin wave spectra\cite{zhao}, were attributed to the ellipticity of the electronic pockets \cite{eremin,tohyama}.

Here, we analyze signatures of this SDW order  in  SI-STM measurements like those reported in Ref.~\onlinecite{chuang}. We do so within the framework of QPI developed in the context of the cuprates \cite{campuzano,vekhter}.
We show that the evolution of the QPI peaks is largely due to  quasiparticle scattering between the electronic bands involved in the SDW formation. Because of the ellipticity of the electron pockets, and due to the fact that only one of the electron pockets is involved in the SDW, the resulting QPI pattern has a pronounced one-dimensional structure. This is in good agreement with the abovementioned experiments\cite{chuang}. Our theory predicts a crossover to two-dimensionality in the QPI to occur at a scale set by (twice) the
SDW gap, $2\Delta_{1}$, which we estimate to be around  $90$meV. The exact value, however, depends on the size of the magnetic moments which can vary from compound to compound.

This paper is organized as follows. We set the stage by introducing model, notation and parameters to describe the underlying electronic structure. We next provide a qualitative account of the gross features of  SI-STM measurements based on the electronic structure of the SDW phase. This we then back up with a detailed calculation of the relevant Green functions within the T-matrix formalism. We close with the discussion of the interplay of inter- and intra-band impurity scattering.

{\it The model.}
We employ an effective mean-field four band model with two circular hole
pockets at $(0,0)$  ($\alpha$-bands) and two elliptic electron pockets at
${\bf Q}_1$ and ${\bf Q}_2$ ($\beta$-bands)\cite{eremin}:
\begin{eqnarray}
\label{eqH}
H_c  & = &  \sum_{\mathbf{k}, \sigma,
i=\alpha_1,\alpha_2,\beta_1,\beta_2} \varepsilon^{i}_{\mathbf{k}} c_{i \mathbf{k}  \sigma}^\dag c_{i \mathbf{k} \sigma} + \nonumber\\
&&  \sum_{{\bf k} \sigma}
\Delta_1 \sigma
\left[ c^\dag_{\alpha_1 {\bf  k} \sigma} c_{\beta_1 {\bf k}+{\bf Q}_1 \sigma}+ H.c.\right]
\end{eqnarray}
We set the  dispersions to
 $\varepsilon^{\alpha_i}_{\mathbf{p}} = t_\alpha\left( \cos p_x +\cos p_y \right) -\mu_i$ and
$\varepsilon^{\beta_1}_{\mathbf{p}}= \epsilon_0 + t_\beta\left( \left[1+\epsilon \right]\cos(p_x+\pi)+\left[ 1-\epsilon \right]\cos(p_y)\right) -\mu_1$,
 $\varepsilon^{\beta_2}_{\mathbf{p}}= \epsilon_0 +
t_\beta\left( \left[1-\epsilon \right]\cos(p_x)+\left[ 1+\epsilon \right]\cos(p_y+\pi) \right)-\mu_1$.  $\epsilon$ accounts for the ellipticity of the electron pockets. Following our previous analysis of the spin wave excitations, we use Fermi velocities and size of the Fermi pockets based on Refs.~\cite{LDA}, namely $t_\alpha=0.85 eV$, $t_\beta=-0.68 eV$,
 $\mu_1=1.54 eV$, $\mu_2=1.44 eV$, $\epsilon_0 = 0.31 eV$, and $\epsilon=0.5$.
For these values, the Fermi velocities are $0.5eV a$ for the $\alpha_1$-band, where $a$ is the $Fe-Fe$ lattice spacing,
 and $v_x=0.27 eV a$ and $v_y=0.49 eV a$ along  $x$- and $y$-directions for
 the $\beta_1$-band, and vice versa for $\beta_2$. We use $a_x = a_y =a=1$.

We introduce the experimentally observed $(\pi,0)$ SDW order parameter,
%(ferromagnetic along one direction and antiferromagnetic along the other),
 within a
standard mean-field approximation:
 ${\vec \Delta}_1 \propto \sum_{\bf p} \langle c^\dag_{\alpha_1\mathbf{p}\delta} c_{\beta_1 \mathbf{p+Q_1} \gamma} \vec{\sigma}_{\delta \gamma} \rangle$. In this state, one of the $\alpha$ fermions couples with only one band of $\beta$ fermions, leaving the other hole and electron bands -- and hence their electron and hole FSs -- unaffected by the SDW. Without loss of generality we direct $\vec{\Delta}_1$ along the $z$- quantization axis.

{\it Results.}
In the following we focus on the discussion of the QPI in the SDW state at energies below twice the SDW gap. Throughout the paper we set $\Delta_1 \approx 45$meV, from a previous analysis
of the experimental data\cite{eremin}.
\begin{figure}[t]
\centering
 \includegraphics[width=1.0\linewidth]{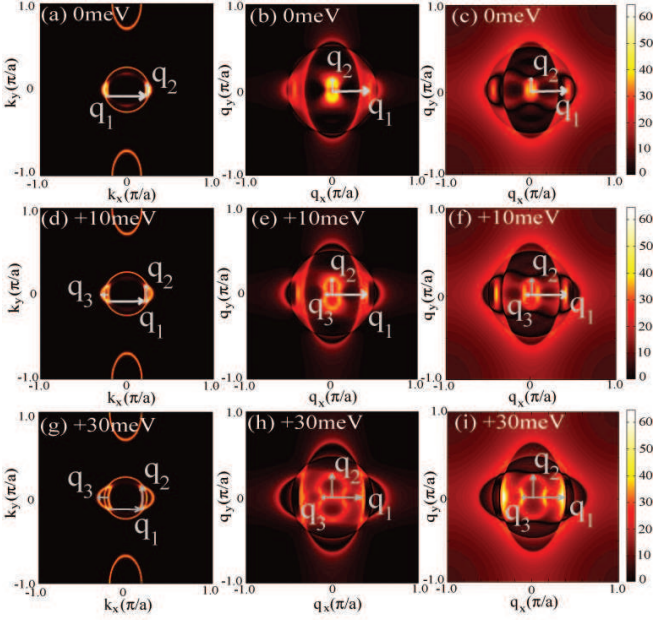}
\caption{(color online) Constant energy intensity maps of the spectral density, $\sum_{\sigma} \mbox{Im}\, \mbox{Tr}\, G_{0\sigma}({\bf k},\omega)$, (left panel) and absolute value of the QPI, $\sum_{{\bf k}\sigma} \mbox{Im}\, \mbox{Tr} \left[G_{0\sigma}({\bf k},\omega) t_{\sigma}({\bf k}, {\bf k+q},\omega) G_{0\sigma}({\bf k+q},\omega)\right]$ for non-magnetic (middle panel) and magnetic (right panel) impurities obtained as described in the text. The arrows denote the characteristic scattering wave vectors which appear in the SDW state. The color bars refer to the intensity in
units of states/eV.}
\label{fig1}
\end{figure}

The simplest way to understand QPI qualitatively is to consider the evolution of the spectral function in the SDW state in the fashion pioneered for the cuprates\cite{campuzano}.
In Fig.\ref{fig1}(a), (d), (g) we show constant energy scans of the spectral density for positive energies. Zero energy, Fig.\ref{fig1}(a), corresponds to the SDW-state Fermi surface (FS).
Its C$_2$ symmetry, lower than the symmetry of the normal state Fermi surface, is immediately apparent. Basically, the FS consists of one hole pocket around the $\Gamma$-point, elliptic electron pockets at $(\pm \pi,0)$ both not involved in the SDW, and two small electron pockets around the $\Gamma$-point that arise due to the folding of one hole and the other elliptic pocket at ${\bf Q}_1$.

It then seems natural to expect that the anisotropy of the QPI will arise due to bands involved in the SDW formation, that is, by the inter- and intra-pocket scattering shown by wave vectors {\bf q}$_1$ - {\bf q}$_3$. This scattering will necessarily reflect the $C_2$-symmetry  induced by SDW order. In particular, {\bf q}$_2$ and ${\bf q}_3$ refer to intra-pocket scattering, while {\bf q}$_1$ represents inter-pocket scattering from the edges of boomerangs which have the largest density of states (DOS).  With increasing energy the scattering between the small pockets starts to interfere with that arising from the large electron pocket, Fig.\ref{fig1}(d)-(g). The anisotropy of the spectral density induced by $(0,\pi)$ SDW order persists to larger energies. However, around $\sim 2\Delta_1$, the influence of the SDW gap disappears (not shown) and the four-fold symmetry of the electronic structure is effectively restored.

The actual QPI which is believed to be measured in SI-STS\cite{davis} arises from quasiparticle scattering by perturbations internal to the sample such as non-magnetic or magnetic impurities. In order to back up the above qualitative picture, we therefore perform a standard analysis of such processes based on a T-matrix description\cite{vekhter}. In particular,  we introduce the impurity term in the Hamiltonian
\bea
{\cal H}_{imp}=\sum\limits_{{\bf k} {\bf  k}^\prime i i^\prime\sigma\sigma^\prime}
\left(
V^{i i^\prime}_{ {\bf  k}{\bf  k}^\prime}
\delta_{\sigma\sigma^\prime}+ J^{i i^\prime}_{\sigma\sigma^\prime }{\bf S}\cdot {\bf \sigma_{\sigma\sigma^\prime}}
\right)
c^\dag_{i{\bf  k}\sigma} c_{i^{\prime}{\bf  k}^\prime\sigma}
\eea
where $V^{i i^{\prime}}_{ {\bf  k}{\bf  k}^\prime}$ and  $J^{i i^\prime}_{\sigma\sigma^\prime }$
define the non-magnetic and the magnetic  point-like
interaction term between the electrons in bands $i$ and $i^\prime$, respectively.
In the following we orient the magnetic impurity in the z-direction (we
did not find a dramatic change in our results by using a general orientation of the impurity).
By changing ${\bf k}+{\bf Q}_1$ to ${\bf k}$ in $c_{\beta_{1}{\bf k}+{\bf Q}_{1}\sigma}$ and
defining the new Nambu spinor as
$
\hat{\psi}_{{\bf k}}^{\dagger}=
(c_{\alpha_2 {\bf k}\uparrow }^{\dagger},c_{\alpha_1 {\bf k}\uparrow }^{\dagger},c_{\beta_1{\bf k}\uparrow  }^{\dagger},c_{\beta_2{\bf k}\uparrow  }^{\dagger}, c_{\alpha_2 {\bf k}\downarrow }^{\dagger}, c_{\alpha_1 {\bf k}\downarrow }^{\dagger},c_{\beta_1{\bf k}\downarrow  }^{\dagger},c_{\beta_2{\bf k}\downarrow  }^{\dagger})$
we can write the Hamiltonian as
\begin{eqnarray}
{\cal H} =
\sum\limits_{{\bf k} }
\hat{\psi}_{{\bf k} }^{\dagger}\hat{\beta}_{{\bf k}}
\hat{\psi}_{{\bf k} }
+
\sum\limits_{{\bf k}{\bf  k}^\prime }
\hat{\psi}_{{\bf k} }^{\dagger}\hat{U}_{ {\bf  k}{\bf  k}^\prime}
\hat{\psi}_{{\bf  k}^\prime}
\end{eqnarray}
where by defining $V^{i i}_{ {\bf  k}{\bf  k}^\prime}=\gamma u_0 $; $V^{i i^\prime}_{ {\bf  k}{\bf  k}^\prime}=\gamma u_Q $
and $J^{i i}_{zz}S_z=\gamma^\prime u_0 $; $J^{i i^\prime}_{zz}S_z=\gamma^\prime u_Q $, the matrices
$\hat{\beta}_{{\bf k}}$ and $\hat{U}_{ {\bf  k}{\bf  k}^\prime}$ are  defined as
\begin{figure}[t]
\centering
 \includegraphics[width=1.0\linewidth]{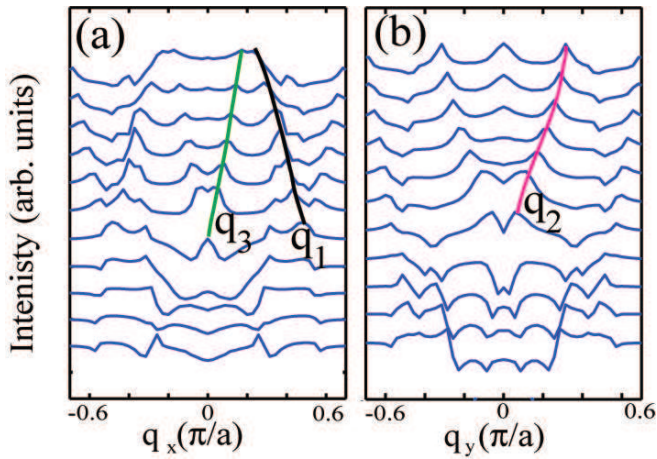}
\caption{(color online) Plots of QPI along AFM, $q_x$, (a) and FM, $q_y$, (b) directions.
The bottom curve is at -50 meV and the top curve is at +50 meV. Consecutive
curves are separated by 10 meV. Red, green and black curves are guides to the eye.}
\label{fig2}
\end{figure}
\begin{eqnarray}
% &&
\hat{\beta}_{{\bf k}}=\left[
\begin{array}{cc}
\hat{\varepsilon}_{ {\bf  k}}^{\uparrow} & 0
\\
0 & \hat{\varepsilon}_{ {\bf  k}}^{\downarrow}
\end{array}
\right];
%
% \nonumber \\&& \hspace{0.25cm}
\hat{U}_{ {\bf  k}{\bf  k}^\prime}=\left[
\begin{array}{cc}
\gamma+\gamma^\prime & 0
\\
0 & \gamma-\gamma^\prime
\end{array}
\right]\otimes
\hat{I}_{ {\bf  k}{\bf  k}^\prime},
\nonumber
\end{eqnarray}
where $\otimes$ is the direct product of matrices and
\begin{eqnarray}
% &&
\hat{\varepsilon}_{ {\bf  k}}^{\sigma} =\left[
\begin{array}{cccc}
\varepsilon^{\alpha_2}_{{\bf k}} & 0 & 0 & 0
\\
0 & \varepsilon^{\alpha_1}_{{\bf k}} & \sigma \Delta_1 & 0
\\
0 & \sigma \Delta_1 & \varepsilon^{\beta_1}_{{\bf k}} & 0
\\
0 & 0 & 0 & \varepsilon^{\beta_2} _{{\bf k}}
\end{array}
\right];
%
% \nonumber \\&& \hspace{0.25cm}
\hat{I}_{ {\bf  k}{\bf  k}^\prime}=\left[
\begin{array}{cccc}
u_0 & u_0 & u_{\bf Q} & u_{\bf Q}
\\u_0 & u_0 & u_{\bf Q}& u_{\bf Q}
\\u_{\bf Q} & u_{\bf Q} & u_0 & u_{\bf Q}
\\u_{\bf Q} & u_{\bf Q} & u_{\bf Q} & u_0
\end{array}
\right].
\nonumber
\end{eqnarray}
Here, we assume that the intraband impurity scattering, $u_0$, is bigger than the interband scattering between the bands separated by a large {\bf Q}, $u_{\bf Q}$, and set $u_{\bf Q} = 0.2 u_{0}$.
  The Green function matrix is obtained via
$G_{ {\bf  k}{\bf  k}^\prime}(\tau)=-\langle T \hat{\psi}_{{\bf k} }(\tau)
\hat{\psi}_{{\bf k}^\prime}^{\dagger}(0)\rangle$,
whence
\bea
% \lefteqn{
G_{ {\bf  k}{\bf  k}^\prime}(\omega_n)
%} && \nonumber\\&&
=G^{0}_{{\bf k}}(\omega_n)[\delta_{ {\bf  k}{\bf  k}^\prime} +
t_{ {\bf  k}{\bf  k}^\prime}(\omega_n)G^{0}_{{\bf k}^\prime}(\omega_n)],
% \nonumber\\
\eea
where $G^{0}_{{\bf k}}(\omega_n)=\left( i\omega_n-\hat{\beta}_{{\bf k}} \right)^{-1}$ is
the bare Green's
function of the conduction electrons. Solving the Dyson equation for the T-matrix
$t_{ {\bf  k}{\bf  k}^\prime}(\omega_n)=\hat{U}_{ {\bf  k}{\bf  k}^\prime}+\sum_{{\bf k}^{\prime \prime}} \hat{U}_{ {\bf  k}{\bf  k}^{\prime \prime}}G^{0}_{{\bf k}^{\prime\prime}}(\omega_n)
 t_{{\bf k}^{\prime \prime}{\bf k}^{\prime}}( \omega_n)$,
the LDOS is obtained via analytic continuation
$i\omega_n\rightarrow E+i 0^{+}$ according to
%\bea
$N^{c}(E,{\bf r}) = -\frac{1}{\pi} \mbox{Im} \, \mbox{Tr} \left[  G(r,r,\omega_n)\right]_{i\omega_n\rightarrow E+i 0^+}$.
%\label{DOS}
%\eea
%
Note that interference between the two partial waves give rise to a spatial modulation of the amplitude of the total wave which is then reflected in the local density of states (LDOS).
We now assume the Born approximation holds and set $u_0 = 0.1t_{\alpha}$.

\begin{figure}[t!]
\centering
 \includegraphics[width=0.8\linewidth]{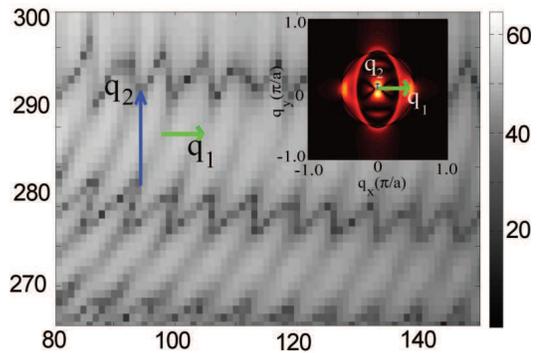}
\caption{(color online) Real space image of the QPI at -7meV away from the impurity site. The inset shows the corresponding QPI image. The QPI interference originating from the structure in the spectral density is reflected in the strong $\sim 10\div 15 a$ modulation along the FM ($x$)-direction and a weak $\sim 5\div 7a$ modulation along the AF ($y$)-direction. In order to model the real Co impurity we adopt the impurity potential as a mixture of the magnetic (20$\%$) and non-magnetic(80$\%$) parts.  Intensity refers to states/eV.
}
\label{fig3}
\end{figure}

We show the absolute value of the resulting QPI for positive energies in Fig.\ref{fig1}(b),(e),(h) for a non-magnetic impurity ($\gamma$=1, $\gamma$'=0)and in
Fig.\ref{fig1}(c),(f),(i) for a magnetic impurity ($\gamma$=0, $\gamma$'=1), respectively. Overall the QPI map resembles well the structure anticipated from the spectral density maps. In particular, we find the overall $C_2$ symmetry of the QPI maps whose structures are determined by scattering at momenta shown on the left panel. In particular, at 10meV the QPI shows peaks at {\bf q}$_2$ and ${\bf q}_3$ which refer to intra-pocket scattering, as well as {\bf q}$_1$  which denote the inter-pocket scattering originating from the edges of the boomerangs.  Overall, the QPI shown in Fig.\ref{fig1}(b), (e) and Fig.\ref{fig3} resembles the one found experimentally in Ref. \cite{chuang}. In particular, we find
that as a result of SDW induced electronic structure reconstruction, the scattering interference modulations are strongly unidirectional and show $C_2$
symmetry with {\bf q}$_1$ and {\bf q}$_3$ induced structure along the AF $q_x$ axis. In the ferromagnetic $q_y$ axis modulations with wave vector {\bf q}$_2$ are also found. As in the electronic structure, this persists only up to a scale set by the size  the SDW gap.

To analyze the energy dispersion of the QPI interference we show in Fig.\ref{fig2} its evolution  along $q_x$ and $q_y$-directions for energies between -50 and 50 meV. The peaks associated with {\bf q}$_2$ along the ferromagnetic $q_x$ direction and ${\bf q}_1$ and ${\bf q}_3$ along the AF $q_y$ direction are dispersive and their velocities are in direct correspondence with the quasiparticle group velocities of the bands involved in the SDW formation.

Finally, in Fig.\ref{fig3} we show a real space image of the QPI away from the impurity position and at -7meV. We observe that the QPI modulation is reflected in periodic structures seen along the $x$ and $y$ directions. In the AF $y$ direction with ${\bf q}_1$ structure, we find a weak 'stripe' pattern with a periodicity of the order of $\approx 7a$.  In addition, we also find a stronger modulation of $10 - 15a$ along the FM $x$ direction, consistent with the structure described as nematic in Ref.~\cite{chuang}. The $x-y$ asymmetry of the structure is again a consequence of the SDW with $(0,\pi)$ magnetic order. We also stress  that these extra modulations arise purely due to scattering between the bands involved in the SDW -- no extra folding of the bands occurs.
\begin{figure}[t!]
\centering
 \includegraphics[width=1.0\linewidth]{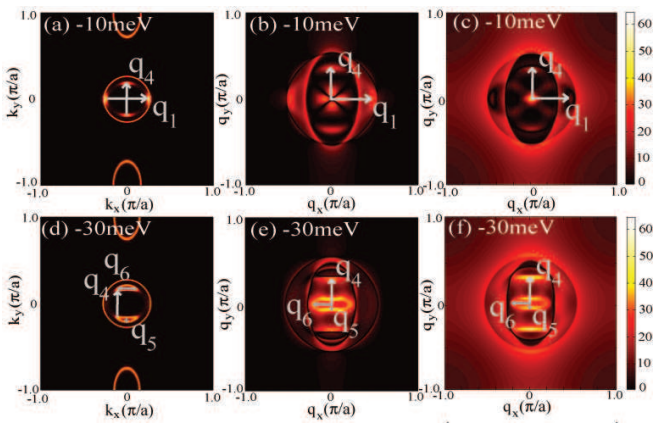}
\caption{(color online) Constant energy intensity maps of the spectral density, $\sum_{\sigma} \mbox{Im}\, \mbox{Tr}\, G_{0\sigma}({\bf k},\omega)$, (left panel) and QPI, $\sum_{{\bf k}\sigma} \mbox{Im}\, \mbox{Tr} \left[G_{0\sigma}({\bf k},\omega) t_{\sigma}({\bf k}, {\bf k+q},\omega) G_{0\sigma}({\bf k+q},\omega)\right]$ for non-magnetic (middle panel) and magnetic (right panel) impurities and negative energies. The arrows denote the characteristic scattering wave vectors which appear in the SDW state. The color bars refer to the intensity in
units of states/eV.
}
\label{fig4}
\end{figure}

One natural question to ask is: How stable is the observed QPI and the resulting real space 'stripe' structure? First, the QPI is sensitive to the size of the SDW gap. For large enough ${\bf \Delta}_1$ the FSs of the bands involved in the SDW are completely gapped and the same is also true for the spectral densities at low energies. In this case, QPI will be  determined by the bands which are not involved in the SDW, and, therefore, its structure will not show strong quasi--1D character. This may explain why the pure AF SDW state does not show any $C_2$-symmetric structure in the parent compounds where the magnetic moment (and the corresponding SDW gap) is quite large. Only when it is reduced upon doping, the bands involved in the SDW are located close to the FS and the QPI structure, described above, becomes visible. Note that the SDW gap used in our calculations corresponds to $\mu\approx 0.4 \mu_B$. An  additional factor why doping might be crucial for an observation of the superstructure  is that the impurities act as pinning centers and the interference is more pronounced once disorder is increased. We also note that as soon as SDW forms due to the nesting of electron and hole bands, the main role of the underlying orbital structure is to modify the absolute intensities of the QPI as this depends on the underlying orbital matrix elements. However, it is unlikely that these factors would restore the four-fold symmetry in the QPI. Instead, it is expected that the corresponding QPI will be the same as in our Fig. 1 but with extra intensity modulation due to orbital matrix elements\cite{remark}.

We note in passing that an increase of the intraband impurity scattering, $u_0$, does not change the results significantly in any way except that at large enough values of $u_0$, one finds local resonances around the impurity site. In addition, the change of the ratio between the interband and intraband scattering is more subtle although the resulting QPI still shows the C$_2$ anisotropy even up to $u_Q=u_0$.
However, there is still another interesting effect that we find upon changing the energy from positive to negative values. In particular, in Fig.\ref{fig4} we show the spectral density and the corresponding QPI from $-10$meV to $-60$meV. At low negative energies, {\it i.e.} above -20meV, the structure remains the same as for 0meV. However, for -30meV we find a rotation of the QPI by 90 degrees. This occurs due to particle-hole asymmetry of the bands contributing to the SDW. In the SDW state, new energies are $E_{{\bf p}}^{c,d}=
\frac{1}{2}\left( \varepsilon^{\alpha_1}_{\bf{p}}+\varepsilon^{\beta_1}_{\bf{p+Q_1}}
\pm \sqrt{(\varepsilon^{\alpha_1}_{\bf{p}}-\varepsilon^{\beta_1}_{\bf{p+Q_1}})^2+4\Delta_1^2} \right)$. While $E^c$ produces small pockets along $x$ (AF) direction which are pronounced for positive energies; $E^d$ yields pockets along the $y$ (FM) direction visible at relatively large negative energies. As a consequence, the QPI rotates  90 degrees at energies lower than -$30$meV. This prediction would be interesting to check experimentally.

In summary, we have presented  a theory for SI-STM measurements in the iron-based superconductor parent compounds in the presence of  $(0,\pi)$ SDW magnetic order. We find that QPI introduced by scalar non-magnetic, as well as magnetic, impurities gives rise to a periodic modulation of the real space LDOS. Because of the ellipticity of the electron pockets, and the fact that only one of the electron pockets is involved in the SDW, the resulting QPI has a pronounced quasi-one-dimensional structure in good agreement with recent SI-STM experiments\cite{chuang}. We further predict that the QPI becomes two-dimensional at energies larger than twice the SDW gap, $2\Delta_{1}$, which as we argue from the analysis of the neutron scattering data should be $\sim 90$meV.

We are thankful to Igor Mazin for sharing with us his results prior to publication and useful comments. I.E. acknowledges the support from the RMES Program (Contract No. N
2.1.1/2985), and NSF (Grant DMR-0456669).


\begin{thebibliography}{99}
\bibitem{remark} In addition, the three-dimensionality of the electronic structure does not change
our results radically due to the significant level of nesting observed in photoemission experiments\cite{prb_new}. Recent calculations also show that the anisotropy of the QPI persists even after averaging over  $q_z$ \cite{mazinprivat}.
\end{thebibliography}
\end{document}